\documentclass[aps,prl,twocolumn,amssymb,showpacs,groupedaddress]{revtex4-1}
\usepackage{graphicx}
\usepackage{xcolor}
\usepackage{amsmath}
\usepackage{float}
\def\beq{\begin{equation}}
\def\eeq{\end{equation}}
\def\bea{\begin{eqnarray}}
\def\eea{\end{eqnarray}}

\begin{document}


\title{Unstable invasion of sedimenting granular suspensions}


\author{Arshad Kudrolli}
\email[]{Corresponding Author: akudrolli@clarku.edu}
\affiliation{Department of Physics, Clark University, Worcester, MA 01610 }
\author{Rausan Jewel}
\affiliation{Department of Physics, Clark University, Worcester, MA 01610 } 
\author{Ram Sudhir Sharma}
\author{Alexander P. Petroff}
\affiliation{Department of Physics, Clark University, Worcester, MA 01610 } 
\date{\today}

\begin{abstract}
We investigate the development of mobility inversion and fingering when a granular suspension is injected radially between horizontal parallel plates of a cell filled with a miscible fluid. While the suspension spreads uniformly when the suspension and the displaced fluid densities are exactly matched, even a small density difference is found to result in a dense granular front which develops fingers with angular spacing that increase with granular volume fraction and decrease with injection rate. We show that the time scale over which the instability develops is given by the volume fraction dependent settling time scale of the grains in the cell. We then show that the mobility inversion and the non-equilibrium Korteweg surface tension due to granular volume fraction gradients determine the number of fingers at the onset of the instability in these miscible suspensions. 
\end{abstract}

\maketitle



Instabilities in the invasion of sedimenting granular suspensions in confined domains are important to natural and industrial systems ranging from microfluidics to hydraulic fracturing, and complementary to fluid flows in porous medium that lead to erosion and rich pattern formation~\cite{Sandnes2011,Bischofberger2014,Kudrolli2016}. Gravitational instabilities because of buoyancy inversion like the Rayleigh-Taylor instability are well known in such suspensions~\cite{Carpen2002,Niebling2010,McLaren2019}. Less obvious are pressure-driven instabilities like the Saffman-Taylor instability~\cite{saffman58,Homsy87,johnsen2008} which arise because of spatial variation in the granular component which affects the effective viscosity of the medium~\cite{krieger}. In the case of a neutrally-buoyant suspension radially invading a fluid confined between two parallel plates, it has been shown that the suspension can break into fingers even when its effective viscosity is greater than the displaced fluid~\cite{tang00}. The meniscus plays an important part in this observation as it blocks the further advance of grains which arrive there in greater proportion because of shear-induced-migration to the faster moving regions away from the boundaries~\cite{kim17}.  The trailing edge of an accumulating annulus of grains is said to become unstable~\cite{tang00,kim17} following a mechanism analogous to viscous fingering observed in Newtonian fluids~\cite{saffman58,paterson81,cardoso95}.
But the phenomena when a meniscus is absent is unclear, as for example when the domain is initially flooded by a similar fluid, and when shear-induced migration is not sufficiently strong to overcome gravity. 

To address Saffman-Taylor-like instabilities in sedimenting suspensions, we discuss experiments with granular suspensions which are injected radially between horizontal plates filled with a similar miscible fluid. Because of the addition of the grains, the viscosity of the injected suspension is effectively higher than the fluid which is displaced. We find that even a small density difference leads the granular component to sediment and lag behind the interface where the fluid components meet and mix, leading to an annular region with lowered mobility. Although the fluids are miscible and the interfacial tension at equilibrium is zero, we show that the non-equilibrium Korteweg surface tension~\cite{Korteweg1901} due to volume fraction gradients plays an equivalent role in determining unstable growth. 



\begin{figure}
\begin{center}
\includegraphics[width=0.45\textwidth]{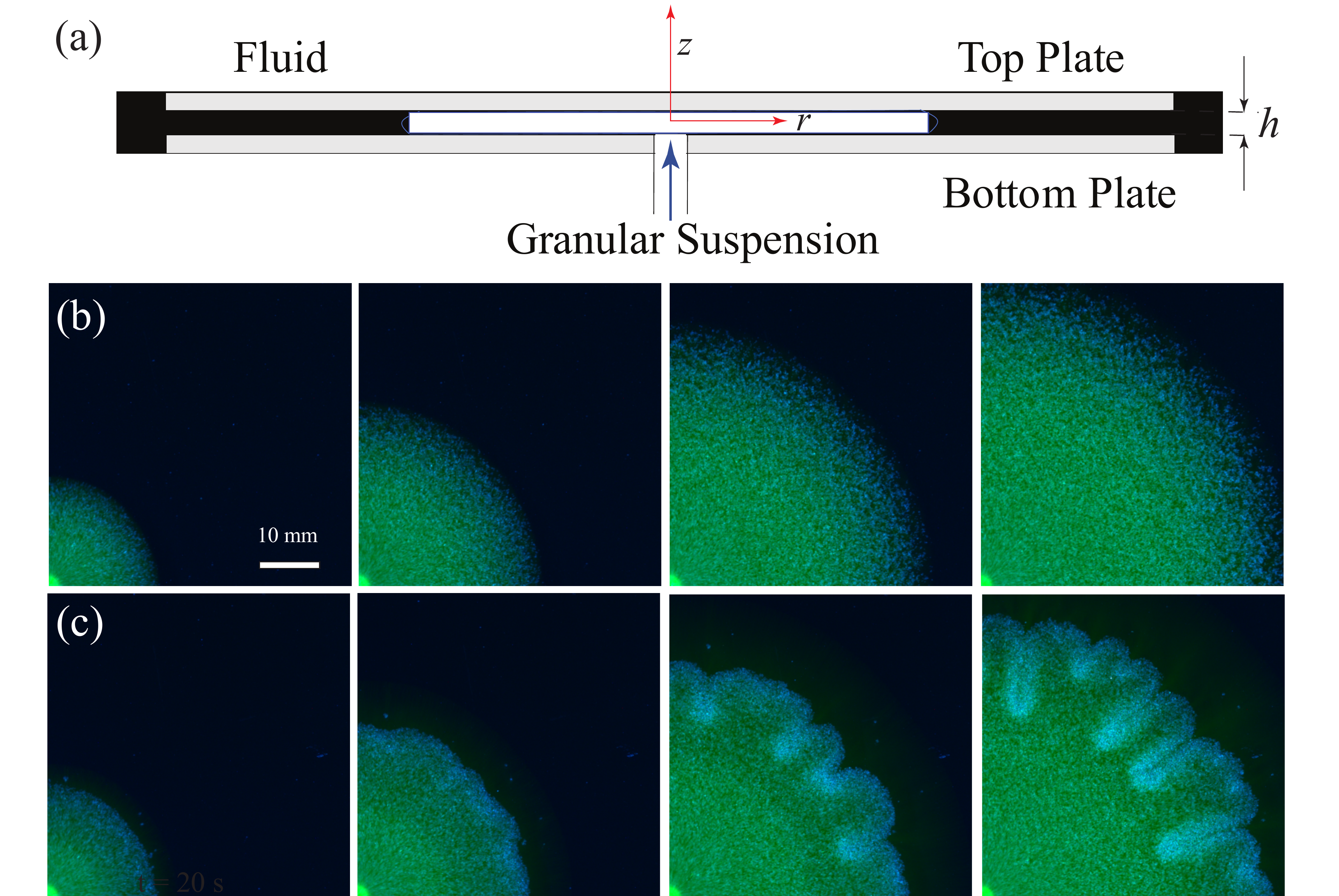}
\end{center}
\caption{(a) A schematic of the experimental cell consisting of parallel plates filled with a miscible fluid and the suspension injection system. Quadrant view of the advancing granular suspension with $\phi_g = 0.2$ and $Q = 0.05$\,cm$^3$\,s$^{-1}$ at time $t = 20$\,s, 50\,s, 100\,s, 150\,s when $\rho_s = \rho_d$ (b), and $\rho_s = 1.07  \rho_d$ (c). The granular component appears blue and the fluid component in the suspension fluoresces green under combined white-blue-UV lighting. Fingers are observed to develop when $\rho_s \neq \rho_d$. See movies in Supplementary Documentation~\cite{supdoc}.  
} \label{fig:schem}
\end{figure} 
Fig.~\ref{fig:schem}(a) shows a schematic of the experimental system
consisting of 20\,cm wide plates separated by distance $h = 1.15$\,mm. A noncohesive granular suspension consisting of sodium chloride, distiled water and polystyrene spheres with diameter $d = 200 \pm 50\,\mu$m and density $\rho_g = 1.07$\,g\,cm$^{-3}$ is prepared with granular volume fraction $\phi_g$ from 0.05 to 0.30~\cite{supdoc}. The density of the fluid is matched to $\rho_g$ by using appropriate salt concentration to ensure that the grains do not sediment and jam inside the injection system~\cite{chopin11,janda15}.  Thus, the density of the injected suspension $\rho_s = 1.07$\,g\,cm$^{-3}$ in all our experiments.  The plates are immersed inside a larger reservoir filled with an aqueous fluid to prevent an air-liquid interface. The density of this fluid $\rho_d$ is varied relative to $\rho_s$ by varying its salt concentration and has a similar viscosity as the fluid component of the suspension $\eta_f \approx 1 \times 10^{-3}$\,Pa\,s at $24^o$C. The suspension is injected through a small hole at the center of the bottom plate with an injection rate $Q$ and the flow can be considered to be in the low Reynolds number regime~\cite{supdoc}. 
We use a cylindrical coordinate system $(r, \theta, z)$ with origin located at the injection point and midway between the top and bottom plates. The system is imaged through the top plate using a megapixel camera. The grains scatter light and appear bright against a dark background with the color of the illumination light which can be mapped to the granular volume fraction $\phi$ at that location~\cite{supdoc}. The fluid in the suspension is visualized by adding a dye which fluoresces green under ultraviolet illumination.


Fig.~\ref{fig:schem}(b) shows snapshots as a suspension with $\phi_g =0.2$ is injected when $\rho_d = \rho_s$. The suspension spreads out uniformly over time with a circular front centered at the injection point. Thus, this experiment shows that the air-fluid meniscus is necessary to observe grain accumulation and fingering in neutrally buoyant suspensions reported previously~\cite{tang00,kim17}. Next, Fig.~\ref{fig:schem}(c) shows an example where $\rho_s = 1.07 \rho_d$. We observe that the suspension initially spreads out in a uniform circle before  fingers develop over time. While in the particular example shown $\rho_s > \rho_d$, we find fingers form just as well when $\rho_s < \rho_d$ over a time scale which only depends on the density differences in the absence of a meniscus.

\begin{figure}
\begin{center}
\includegraphics[width=0.49\textwidth]{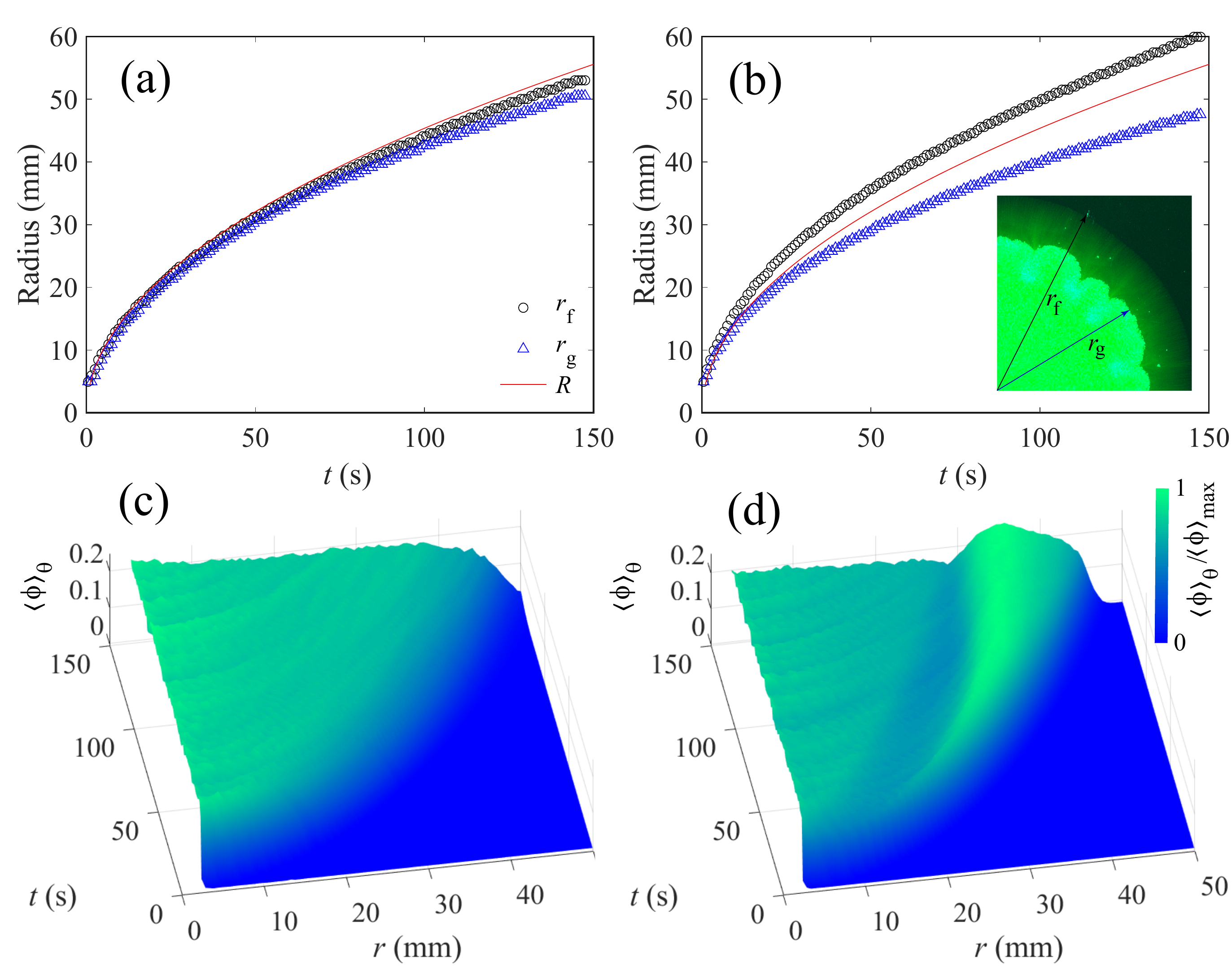}
\end{center}
\caption{(a,b) The radius of the fluid front $r_f$ and granular front $r_g$ as a function of time when $\rho_s = \rho_d$ (a) and  $\rho_s = 1.07 \rho_d$ (b). Inset: A green color enhanced image used to identify $r_f$, which leads $r_g$. The original image corresponds to $t=100$\,s in Fig.~\ref{fig:schem}(c). An estimated front of the suspension $R(t)$ plotted assuming  Eq.~\ref{eq:R} fit $\alpha_h = 1.22$. (c,d) The average volume fraction $\langle \phi \rangle$ as a function of distance $r$ and time $t$ when $\rho_s = \rho_d$ (c) and  $\rho_s \neq \rho_d$ (d). The color map is scaled by the maximum volume fraction $\langle\phi\rangle_{\rm max}=0.25$ in each case to highlight the variation in $\langle \phi \rangle_\theta$.  
}
\label{fig:front}
\end{figure}

We identify the angle-averaged radial distance of the fluid front $r_f$ and the granular front $r_g$ from the injection point by image processing and plot them in Fig.~\ref{fig:front}(a) for $\rho_s = \rho_d$ and Fig.~\ref{fig:front}(b) for $\rho_s \neq \rho_d$. While $r_f$ and $r_g$ are observed to essentially overlap in Fig.~\ref{fig:front}(a) with small differences due to residual differences in their densities, $r_f$ is observed to systematically lead $r_g$ when $\rho_s > \rho_d$ in Fig.~\ref{fig:front}(b), as is also clear from the green color enhanced image shown in the inset.  If the suspension spreads uniformly, then the radius of the suspension increases as 
\begin{equation}
R(t) = \alpha_h \sqrt{Qt/\pi h},
\label{eq:R}
\end{equation} 
where $\alpha_h = 1$ if the suspension and displaced fluid do not overlap, and $\alpha_h > 1$ if they mix or overlap. For example, if the flow is perfectly parabolic with nonslip boundary conditions at the top and bottom, $\alpha_h = \sqrt{2}$. $R(t)$ corresponding to a fitted value of $\alpha_h =1.22$ is observed to well describe the data in Fig.~\ref{fig:front}(a) showing some degree of mixing. The same curve is plotted in Fig.~\ref{fig:front}(b), and is observed to systematically lag $r_f$, and lead $r_g$.  In fact in the case where $\rho_s = 1.07 \rho_d$, we find 
$r_f$ is fitted by Eq.~\ref{eq:R} with $\alpha_h = 1.35$, and $r_g$ is fitted by Eq.~\ref{eq:R} with $\alpha_h = 1.07$. Thus, we find that the advance of the granular front slows down relative to the fluid phase leading to a buildup of $\phi$ near the front as we quantify next.

We obtain the azimuthally average granular volume fraction $\langle \phi \rangle_\theta$ as a function of radial distance $r$ and plot it in Fig.~\ref{fig:front}(c) and Fig.~\ref{fig:front}(d) for $\rho_s = \rho_d$ and $\rho_s \neq \rho_d$, respectively, over the time that the suspension is injected. We observe in Fig.~\ref{fig:front}(c) that $\langle \phi \rangle_\theta$ is constant and equals $\phi_g$ before decreasing monotonically to zero at the front. The effective viscosity of dense suspensions is given by~\cite{krieger}:
\begin{equation}
\eta_s = \eta_f \left(1-{\phi}/{\phi_c}\right)^{-2.5 \phi_c},
\label{eq:kd}
\end{equation}
where $\phi_c \approx 0.6$ is the granular volume fraction at which the suspension jams~\cite{supdoc,Kudrolli2016}. Thus $\eta_s$ monotonically decreases with $r$ and no fingering instability is observed under these conditions~\cite{saffman58,paterson81}. By contrast, 
we observe in Fig.~\ref{fig:front}(d) that $\langle \phi \rangle_\theta$ initially decreases monotonically with $r$, but becomes non-monotonic with the formation of a peak which increases in strength over time. This peak showing the accumulation of grains results in a relative increase in $\eta_s$ before decreasing to $\eta_f$. While the mechanism by which this annulus arises is due to a different reason than in neutrally buoyant suspensions~\cite{kim17}, it nonetheless results in a mobility inversion which gives rise to conditions in which the invasion can become unstable~\cite{saffman58,paterson85}.   

\begin{figure}
\begin{center}
\includegraphics[width=0.5\textwidth]{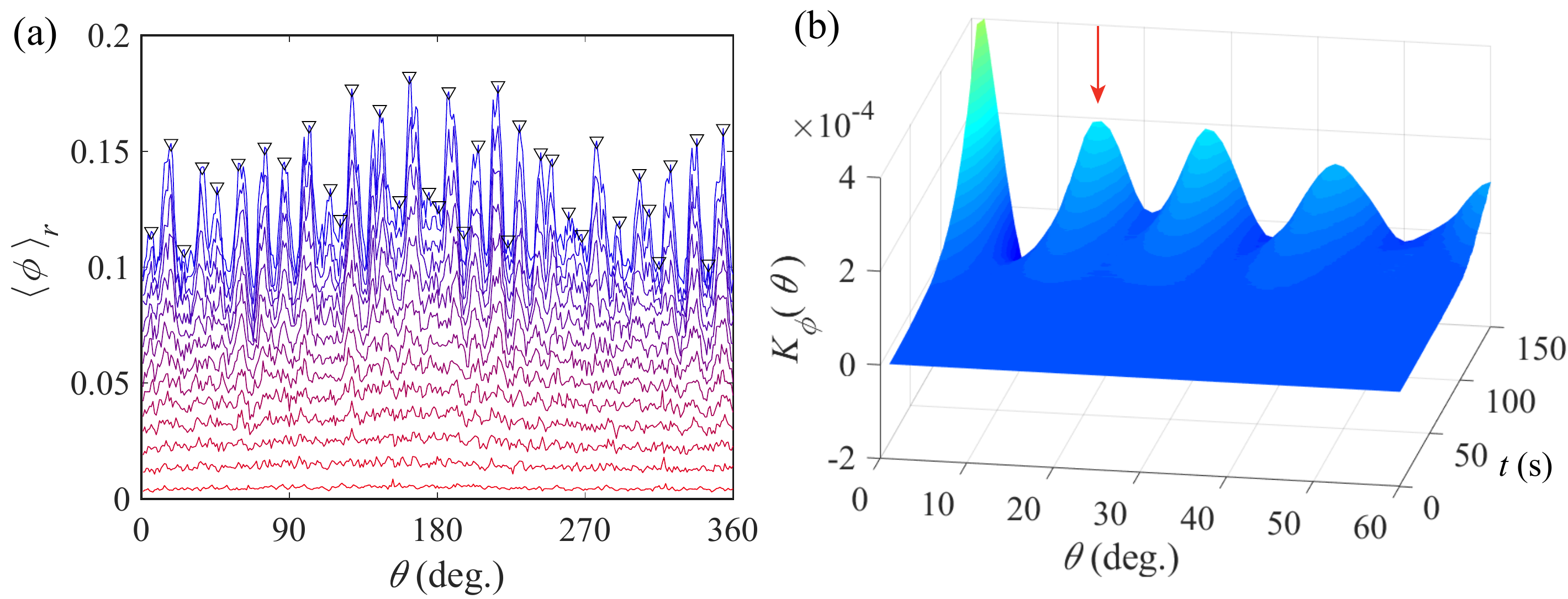}
\end{center}
\caption{(a) Radial cumulative density reveals peaks corresponding to development of fingers denoted by markers plotted in 10\,second time intervals. (b) The angular self-correlation function $K_\phi$ 
when $\rho_s = 1.07 \rho_d$. The peak indicated by the arrow corresponds to  growth of the maximum amplitude mode.  
}
\label{fig:peaks}
\end{figure}
 
To quantify the development of fingers, we obtain the radially averaged volume fraction of the grains $\langle \phi \rangle_r$ in one degree sectors out to a fixed distance $r=7$\,cm. Fig.~\ref{fig:peaks}(a) shows $\langle \phi \rangle_r$ as a function of angle $\theta$ for $\rho_s \neq \rho_d$. It shows that $\langle \phi \rangle_r$ rises more or less uniformly as a function of $\theta$ as the suspension moves out initially, but peaks develop over time as identified by the markers. We calculate the self-correlation function $K_\phi(\theta) = \langle \langle\phi\rangle_r(\theta_o + \theta) \langle\phi\rangle_r(\theta_o) \rangle -  \langle\langle\phi\rangle_r(\theta_o) \rangle^2$ by averaging over $\theta_o$ 
to identify the mean angle between fingers. Fig.~\ref{fig:peaks}(b) shows the corresponding $K_\phi(\theta)$, where we observe a clear emergence of a peak  at $\theta_m = 14$\,deg. $\pm 1$\,deg. at a time $t_c \approx 40$\,s  -- denoted by the arrow. The angle $\theta_m$ where the peak occurs corresponds to the mean angle between fingers and can be associated with the wavelength of the instability $\lambda \approx \theta_m r_g \approx 25$\,mm. Thus $\lambda \gg h$, and much larger than buoyancy-driven rolls observed in Newtonian fluids with characteristic length scale given by $h$~\cite{haudin14}.



\begin{figure}
\begin{center}
\includegraphics[width=0.45\textwidth]{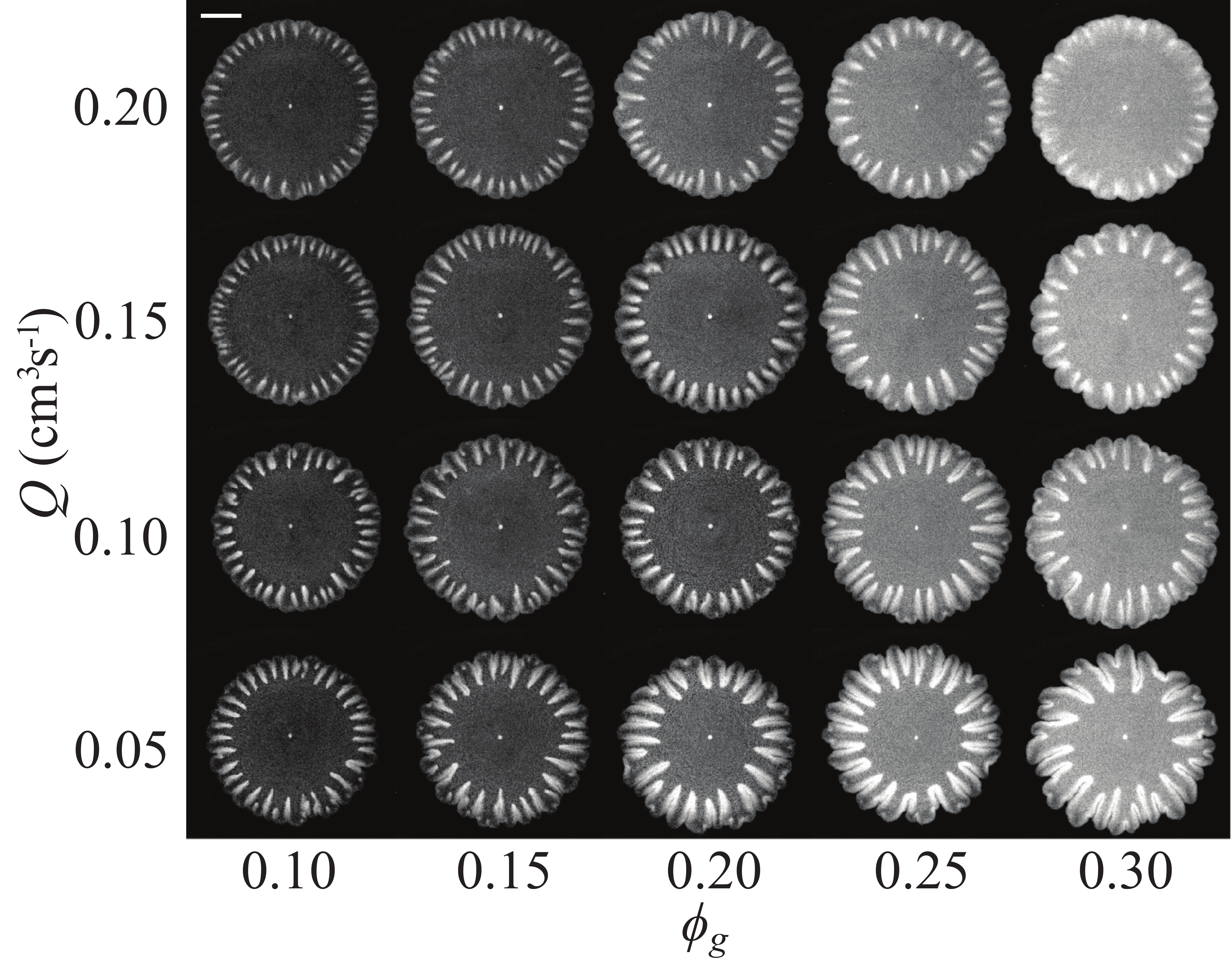}
\end{center}
\caption{Various finger patterns observed with a  B\&W camera after a fixed volume of suspension 
is injected ($\rho_s = 1.07 \rho_d$). The scale bar is 20\,mm.
}
\label{fig:table}
\end{figure}

\begin{figure}
\begin{center}
\includegraphics[width=0.5\textwidth]{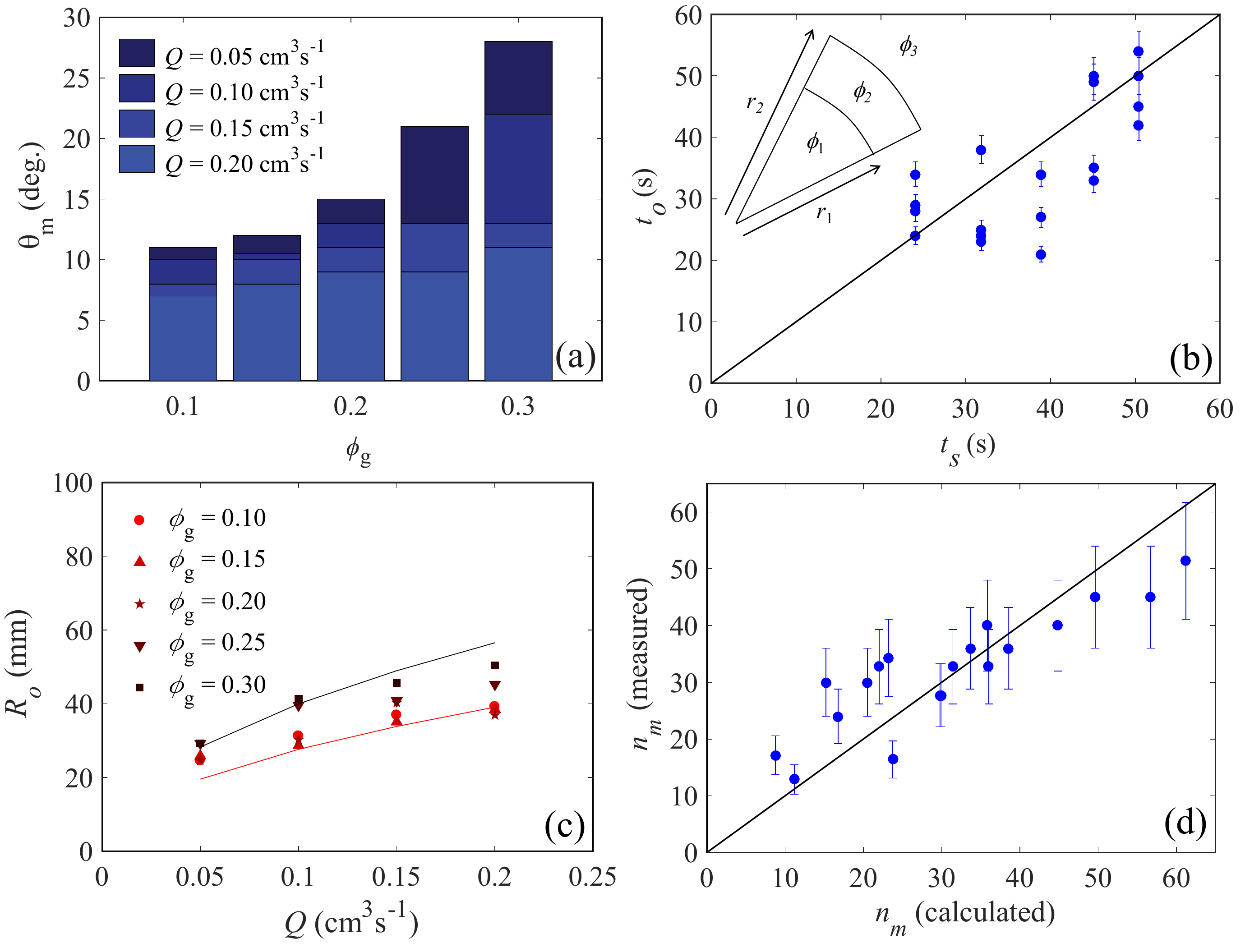}
\end{center}
\caption{(a) Systematic variations in $\theta_m$ can be observed with $Q$ and $\phi_g$. (b) The observed time when annulus form $t_o$ versus calculated settling time $t_s$ of the grains solving Eq.~\ref{eq:settle} under various $Q$ and $\phi_g$. The line corresponds to $t_o/t_s = 1$. (c) The measured radius $R_c$ when fingers appear. The lines correspond to calculated $t_s$ obtained by solving Eq.~\ref{eq:settle}.  (d) The calculated mode number from Eq.~\ref{eq:nmt} is found to capture the measured mode number $n_m$ over the entire range of the $Q$ and $\phi_g$. The line corresponds to slope 1.}
\label{fig:toModel}
\end{figure}

We varied  $\phi_g$ and $Q$ to investigate $\theta_m$ and understand the emergence of the dominant mode. Fig.~\ref{fig:table} shows the fingering patterns observed after a fixed volume of the suspension $V_s= 20$\,cm$^3$ is injected in each trial. No dye is added to the liquid, and the brighter regions correspond to the granular component. We observe that the size of the final patterns is roughly the same in all the trials, but the number of fingers, and where they start, is observed to vary with $\phi_g$ and $Q$. Fig.~\ref{fig:toModel}(a) shows measured $\theta_m$ when a clear peak emerges in $K_\phi(\theta)$ corresponding to the onset of fingering instability. We observe $\theta_m$ increases systematically with $\phi_g$ and decreases systematically with $Q$. Based on these observed characteristics of the fingers, we next develop an understanding of when and where the instability occurs.  

We consider an annulus with inner radius $r_1$ and outer radius $r_2$ where grains begin to accumulate as sketched in the inset to Fig.~\ref{fig:toModel}(b). We assume the volume fraction in the inner region  $\phi_1 \approx \phi_g$, the volume fraction in the annular region $\phi_2 \approx \phi_{max}$, and the volume fraction in the outer region $\phi_3 \approx 0$. 
We observe that $\phi_{max}$ remains close to $\phi_g$ even as the fingers develop~\cite{supdoc}, where friction between the grains and the substrate can be important~\cite{dumazer16}. Thus, $\phi$ does not have to reach $\phi_c$ for fingering to occur, although such values can be reached later in the development of the fingers. We estimate the time over which this annulus develops by considering the time scale $t_s$ determined by the settling speed of
the grains. The Stokes settling speed is given by
$v_s  = \frac{2}{9} f \frac{a^2 g \Delta \rho}{\eta_f}$,
where $f$ is the hindrance due to the presence of the other grains,
and is given by $f =
{\eta_f(1-\phi_g)}/{\eta_s(\phi_g)}$~\cite{leighton86}, and $\Delta \rho \approx (\rho_s - \rho_f)/2$. 
To find $t_s$, we consider a column of water
mixed with the grains. Initially, the volume fraction of grains is that
of the injected suspension, $\phi_g$ and the height of the column is the
distance between the plates, $h$. The height of the column decreases
as grains settle, and from
conservation of mass, the average volume fraction of grains in the
collapsing column $\phi_z$ is a solution of~\cite{supdoc},
\begin{equation}
\dfrac{\mathrm{d}^2\phi_z}{\mathrm{d}t^2}=\left(\dfrac{2}{\phi_z}+\dfrac{1}{v_s}\dfrac{\mathrm{d}v_s}{\mathrm{d}\phi_z} \right)\left(\dfrac{\mathrm{d}\phi_z}{\mathrm{d}t}\right)^2.
\label{eq:settle}
\end{equation}
We integrate Eq.~\ref{eq:settle} numerically with the boundary condition that $\phi_z(0) = \phi_g$ at $t=0$ to find $t_s$ at
which $\phi_z$ is within $5\%$ of $\phi_c$. We then plot $t_s$ as a function of
the observed time $t_o$ for the onset of dense granular annulus in
Fig.~\ref{fig:toModel}(b), and observe a good correspondence. 

Once the granular components settle, they slow down because of non-slip boundary conditions at the bottom surface, leading to an effective decrease of the thickness of the flowing region between the plates. This leads to a decrease in mobility because it is inversely proportional to the square of the thickness of the flowing region between the plates. Thus, the radius  $R_o$, where the instability occurs, can be estimated using  
Eq.~\ref{eq:R} as $R_o \approx \alpha_h \sqrt{Qt_s/\pi h}$. Plotting $R_o$ for $\phi_g = 0.1$ and 0.3 in Fig.~\ref{fig:toModel}(c), we note that it describes well the measured radius at which the granular annulus form for $\phi_g$, with intermediate cases in between these two limits.    

After mobility inversion occurs due to formation of the annulus, perturbations of the front can grow provided they can overcome surface tension.
While an analysis of unstable modes in miscible fluids notes that all modes are unstable, with a lower cutoff given by the thickness of the cell~\cite{paterson85}, this is clearly not what we observe.  Thus, we examine the calculated number of fingers at radial distance $r$ from the injection point in the case of fluids with interfacial tension $\Gamma$~\cite{paterson81}:  
\begin{equation}
n =  \sqrt{\frac{1}{3} [\frac{6 Q r (\eta_2 - \eta_1)}{\pi h^2 \Gamma}] +1} \,,
\end{equation}
where, $\eta_1$ and $\eta_2$ are the viscosities of the inner and outer fluids respectively. While the surface tension between two miscible fluids is zero at equilibrium, the presence of volume fraction gradients at the front implies that an off-equilibrium  Korteweg surface tension can exist at the interface~\cite{Korteweg1901,Truzzolillo2014,Chui2015}, and in the case of suspensions is given by~\cite{Truzzolillo2014},
$\Gamma_e = \frac{\kappa}{\delta} \Delta \phi^2$,
where $\kappa$ is the Korteweg constant, and $\delta$ is the radial distance over which the volume fraction changes by $\Delta \phi$. In case of thermal systems $\kappa$ is proportional to the temperature~\cite{Truzzolillo2014}. Because granular suspensions are athermal, a granular temperature set by the local shear rate may play a similar role~\cite{Brennen2005}. 

For sufficiently small $|r_2 - r_1|$ and incompressible fluids, any perturbation of the interface between $\phi_1$ and $\phi_2$, results in a deformation of interface between $\phi_2$ and $\phi_3$. Thus, while the change in volume fraction is small between the interface 1 and 2, the surface tension effect is essentially dominated by the value going from interface 2 to 3. Then, $\Gamma_e = \frac{\kappa}{\delta} \phi_{max}^2$.  Because $\phi_{max} \approx \phi_g$ as the fingers develop~\cite{supdoc}, we  linearize the change in viscosity in terms of the change of volume as $(\eta_2 - \eta_1) \approx 2.5 \eta_f (\phi_2 - \phi_1)$. 

Then,  we have the mode $n_m$ with maximum amplitude, assuming 
$n_m \gg 1$, as 
\begin{equation}
n_m \approx \alpha_m \sqrt{\frac{5 \alpha_h}{\pi^{3/2}}} \sqrt{ \frac{\delta}{\kappa}}\sqrt{ \frac{\eta_f {t_s^{1/2} Q^{3/2}}  (\phi_2 - \phi_1)}{ h^{5/2}  \phi_{max}^2}}\,,
\label{eq:nmt}
\end{equation}
where, $\alpha_m \simeq 0.422$~\cite{Truzzolillo2014} is introduced to account for the fact that we identify the mode with the maximum amplitude. It is further possible to approximate $(\phi_2 - \phi_1)$ with $(\phi_{max} - \phi_g)$. However, because of the sensitivity of $n_m$ on this difference, we have calculated $n_m$ based on the actual measured difference rather than this last approximation.   
Fig.~\ref{fig:toModel}(d) shows the comparison of the measured number of fingers $n_m = 360/\theta_m$ versus those calculated using Eq.~\ref{eq:nmt}. We observe that the data is well described by a line with $\kappa/\delta = 0.017$\,N\,m$^{-1}$ and goodness of fit R$^2 = 0.73$~\cite{supdoc}. Thus, we understand the instability occurs when a mobility inversion develops due to increasing $\phi$ at the front, with mode number given by the Korteweg surface tension.   

In conclusion, we have demonstrated that the invasion of a suspension into a miscible fluid is unstable when their densities are not exactly matched. We show that granular sedimentation in the resulting mixed suspension gives rise to mobility inversion and a Saffman-Taylor-like instability even in the absence of a meniscus. We then explain the observed maximum amplitude mode as a function of system parameters by noting the contribution of a Korteweg-like  interfacial tension on the stability of the patterns in miscible suspensions. The relative contribution of granular temperature in estimating this non-equilibrium tension remains an interesting avenue for future research.

\begin{acknowledgments}
We thank Benjamin Allen for discussions. This work was supported by DOE DE-SC0010274. This work was also partially supported by the National Science Foundation under Grant No. CBET 1805398.
\end{acknowledgments}

\bibliographystyle{unsrt}

\end{document}